# Dynamics of magnetization and carriers at the onset of the photo-excited precession of magnetization in (Ga,Mn)As

T. Matsuda, K. Nishibayashi and H. Munekata

*Imaging Science and Engineering Laboratory, Tokyo Institute of Technology, Japan*

4259-J3-15 Nagatsuta, Midori-ku, Yokohama 226-8503, Japan

**Abstract**

Photo-excited precession of magnetization in (Ga,Mn)As is investigated by measuring time-resolved magneto-optical response and transient differential reflectivity with pump-and-probe technique. In the time region less than 1 ps, rapidly oscillating and spike-like signals are observed, respectively, with excitation of below and above the GaAs band gap. Analysis with gyromagnetic model and autocorrelation function concludes that those signals are not attributed to ultrafast demagnetization but due to interference between pump and probe pulses incorporating sub-ps carrier dynamics characteristic of low-temperature grown semiconductors. Photo-ionization of Mn ions ($Mn^{2+} \rightarrow Mn^{3+}$) is proposed as a mechanism which dynamically induces orbital angular momentum and affects hole-mediated magnetic anisotropy in (Ga,Mn)As.

Ultrafast magnetization dynamics induced by femtosecond laser pulses of visible and near infrared light has addressed a question to what extent excited electrons can influence ordered spins. In metals, laser-excited electrons thermalize rapidly in the range of 0 - 5 ps through intra- and inter-band scattering, during which spin ordering is temporarily lost [1-3]. This phenomenon, ultrafast demagnetization, has been found to trigger precession of the recovered magnetization with an external magnetic field applied not parallel to the easy axis [2-4]. In insulators, the band gap intercepts thermalization of laser-excited electrons, which significantly reduces the amount of energy released within the range of 5 ps. Naturally, non-thermal phenomena, such as transient photo-magnetization due to the direct excitation of magnetic ions [5,6] and magnetic polaron formation [7,8], and photo-excited precession of magnetization (PEPM) caused by the inverse Faraday effect with virtual excitation [3,9,10], become important in magnetization dynamics.

In the present work, we are concerned with PEPM in (Ga,Mn)As [11] in which ferromagnetic order appears when Mn local moments and holes coexist with concentration of $10^{20}$ cm$^{-3}$ or higher [12], a regime in between insulators and metals.

Despite numbers of works on PEPM [13-16], however, experimental evidence revealing the mechanism of triggering the precession has not been presented up to now. Moreover, in view of metal vs. insulator, majority of works dealing with hole-mediated ferromagnetism in (III,Mn)V semiconductors have treated holes in the valence band as a primary medium to control ferromagnetism [17-20] on the basis of the *p-d* Zener model [21,22], whereas studies on controlling ordered spins through direct manipulation of Mn ions via hidden states have been scarce.

We approach to those problems by studying the dynamics of magnetization and carriers at the onset of PEPM in the time region less than 1 ps under weak and moderate excitation conditions (0.1 – 10 $\mu$J/cm$^2$ per pulse) with different excitation photon energies. We report that the excitation band for PEPM not only overlaps with that of a GaAs host but also extends inside the gap for about 100 meV, and that the PEPM is *not accompanied* by ultrafast demagnetization. We further discuss, on the basis of the observed below-gap PEPM and reflectivity data, that PEPM is triggered by the photo-ionization of acceptor Mn ions (Mn$^{2+}$ → Mn$^{3+}$) through which magnetic anisotropy is dynamically rotated.

The sample studied in this work is a 100 nm-thick $Ga_{0.98}Mn_{0.02}As$ epitaxial layer grown on a GaAs/GaAs(001) substrate by molecular beam epitaxy at the substrate temperature of 235 °C [14]. The Curie temperature is around 45 K, and predominant magnetization easy axis lies along the in-plane, GaAs ⟨100⟩ axis at low temperatures (< 20 K), as confirmed by magnetization measurement. A resistance maximum has appeared around the Curie temperature in a resistance–temperature curve (Fig. 1(a)), showing that the sample is in so called metallic regime [12].

A one-color pump-and-probe (P-P) system based on a mode-locked Ti:sapphire laser was utilized to measure time-resolved magneto-optical (MO) and time-resolved differential reflectivity (ΔR/R; DR) signals. Duration and repetition rate of the laser were around 150 fs, and 76 MHz, respectively. Wavelength of the laser was varied between 750 and 900 nm ($h\nu$ = 1.38 to 1.65 eV). Pump and probe pulses, whose polarizations were both adjusted parallel with GaAs [010] axis, were focused into the same spot with the diameter of 100 $\mu$m using a single lens. The incident angles were around 6 and 3° for pump and probe pulses, respectively. MO signals were composed of polar Kerr rotation (PKR) and magnetic linear birefringence (MLB) [23] which represent changes in the magnetization component along the out-of-plane and in-plane directions, respectively. Time resolution of the measurement in the ultrafast time region (~ 4 ps) is 26 fs which is determined by the precision of a mechanical stage in the P-P system. All P-P measurements were carried out at 10 K.

Shown in Fig. 1(b) are long-term temporal profiles of MO signals obtained for six different P-P photon energies. The fluence of pump and probe pulses were kept constant at $I_{pump}$ = 1.7 $\mu$J/cm$^2$ and $I_{probe}$ = 84 nJ/cm$^2$, respectively. The MO profile obtained with $h\nu$ = 1.57 eV consists of a relatively slow raising/declining component (a dashed line) and a damping oscillation component (a solid line), added with a spike-like component (surrounded by a rectangle) in the time range of a few ps. Components represented by dashed and solid lines have been reported previously, and assigned, respectively, as signals reflecting the rotation of the effective magnetic field $H_{eff}$ toward the out-of-plane direction and the precession motion of magnetization induced by the rotated $H_{eff}$ [14]. The mechanism of this process has been hypothesized in terms of a change in hole-mediated magnetic anisotropy [21, 22] due to repopulation and annihilation of photo-generated holes in the valence band.

At the P-P photon energies of $h\nu$ = 1.51 eV, a large, long-lived exponential component with primary time constant of around 2300 ps appears. The oscillatory component is clearly noticeable, but is accompanied by the phase shift toward shorter time region. A trace of the spike-like component is still observable. Further decreasing the P-P photon energy results in complete disappearance of the large, long-lived component, and only damping oscillation component is observed, as seen in the profile obtained with $h\nu$ = 1.48 eV. The amplitude of oscillations decreases gradually with reducing the P-P energies, and PEPM diminishes at $h\nu$ = 1.38 eV ($\lambda$ = 900 nm). Occurrence of PEPM with P-P energies below the GaAs band gap was reported earlier [15], but critical discussion was not made in view of the fundamental mechanism of the PEPM. Taking into account of a sharp fundamental absorption edge of GaAs, the large exponential response observed at $h\nu$ = 1.51 eV is inferred to be due to carrier dynamics in a substrate [24], and we will not discuss this any further in the present manuscript.

Throughout the entire photon energy range studied in this work, no noticeable dependence of oscillation signals on pump light polarization has been observed. This fact indicates that inverse Faraday [9,10] and inverse Cotton-Mouton effects [25] are not responsible for the observed PEPM. The observed phase shift in the oscillation component with the P-P energies of $h\nu \geq 1.48$ eV (Fig. 1(b)) is attributed to the delay in the onset of PEPM caused by photo-generated carriers near/above the valence band edge and time dependent change in MO coefficients. Those effects will be discussed in detail in a separate paper in relation to the mechanism of PEPM with the above-gap excitation [26].

Shown in Fig. 1(c) are normalized MO and DR temporal profiles for $h\nu = 1.57$ eV in the ultrashort time range between $-2$ and 4 ps with the pump fluence of $I_{pump} = 1.7$ $\mu$J/cm$^2$. Both profiles exhibit spike-like components which superimpose fairly well to each other, indicating ultrafast changes in the diagonal component of complex optical index. Contribution of two different mechanisms can be inferred: a slight change in band structures due to ultrafast demagnetization, or a slight, incoherent change in optical index due to ultrafast relaxation of photo-generated carriers [27].

In Fig. 1(d), temporal MO and DR profiles in the ultrashort time region are shown for $h\nu = 1.44$ eV ($I_{pump} = 1.7$ $\mu$J/cm$^2$). The spike-like component totally disappears, despite the fact that PEPM is still active (Fig. 1(b)). Only detected are rapidly oscillating signals around time zero region, which resembles the interference between pump and probe pulses incorporating the coherent photo-induced change in refractive index. Those data suggest two important points; firstly, ultrafast demagnetization is not a prerequisite for PEPM in (Ga,Mn)As, and secondly, the below-gap excitation does not yield an incoherent variation of optical index which usually takes place when carriers are photo-generated and scattered. [28]

To further examine the experimental data quantitatively, we have carried out two types of model calculation. The first one is of a magnetic origin based on the Landau-Lifshitz-Bloch (LLB) equation, and the second one is of a non-magnetic origin based on autocorrelation function of pump and probe pulses convoluted with the ultrafast response of refractive index due to photo-generated carriers [27,29].

A change in the magnitude of magnetization enters in the third term of the LLB equation [30,31], as represented by Eq. 1.

$$\frac{\partial M}{\partial t} = \gamma M \times H + \gamma M_s \frac{\alpha_\perp}{M^2} M \times (M \times H) - \gamma M_s \frac{\alpha_{//}}{M^2}(M \cdot H)M \qquad (1)$$

Here, $\gamma$ is the gyromagnetic constant, $M_s$ saturation magnetization, $H$ the effective magnetic field, whereas $\alpha_\perp \sim 0.25$ [14] and $\alpha_{//}$ are dimensionless transverse and longitudinal damping factors, respectively. Ultrafast demagnetization is expressed by the exponential function $\alpha_{//}(t) = \alpha_{//}^0 \exp(-t/\tau_{dem})$ with its magnitude $\alpha_{//}^0 = 0.09$ and 0.15 for $h\nu = 1.44$ and 1.57 eV, respectively. The lifetime $\tau_{dem} \sim 0.4$ ps is chosen in view of reproducing the experimental data. A protocol of calculation is same as that in Ref. 14, and $\boldsymbol{H}$ is assumed $\boldsymbol{H} = \boldsymbol{H}_0 + \boldsymbol{H}_{ext} + \boldsymbol{H}_{dem}$ in which the in-plane crystal anisotropy field $\boldsymbol{H}_0 \sim 2000$ Oe, an external field $\boldsymbol{H}_{ext} = 0$, and the demagnetizing field $\boldsymbol{H}_{dem} = 0$. The calculated MO profiles exhibit a *negative* spike-like component followed by a straight, horizontal line, as shown by open circles in Figs. 1(c), (d) and Fig. 2(a). This is reasonable since, firstly, demagnetization always results in the reduction in the initial in-plane magnetization ($M_x$), and, secondly, there is no mechanism which connects $\boldsymbol{H}$ and the ultrafast demagnetization in the Eq. (1). In detail, our calculation yields polarization rotation of 3.7 and 1.4 $\mu$rad, for $h\nu = 1.57$ and 1.44 eV, respectively. Those values represent a reduction in magnetization of 0.19 (Fig. 1(c)) and 0.13 % (Fig. 1(d)), respectively.

If we intentionally introduce in the Eq. (1) the *out-of-plane* rotation of the effective magnetic field $\boldsymbol{H} = \boldsymbol{H}_{eff} = H_0 \, (\cos\theta_{eff}, 0, \sin\theta_{eff})$ in addition with ultrafast demagnetization contribution, we obtain both spike-like and oscillation components, as represented by the dashed line in Fig. 2(a). Here, we use $\theta_{eff} = \theta_{eff}^0 \exp(-t/\tau_1)\{1-\exp(-t/\tau_2)\}$ with $H_0$ = 1800 Oe, $\tau_1$ = 90 ps, $\tau_2$ = 130 ps and constant MO coefficient with the ratios $I_y / I_x$ = 1 and $I_z / I_x$ = 1.71 for calculation. Let us note that calculations with positive MO coefficients yield positive PEPM signals and negative ultrafast demagnetization signals (Fig. 2(a)), as expected. The discrepancy appearing between the simulated and experimental data in the time region longer than 50 ps (Fig. 2(a)) is due to the damping term in the LLB equation which is only valid for the system with a small damping factor ($\alpha_\perp$ < 0.01). Simulation with Landau-Lifshitz-Gilbert equation using same parameters for the LLB calculation yields a reasonable fit to experimental data (dots in Fig. 2(a)) [26]. As a whole, the spike-like component observed experimentally in MO profiles can not be reproduced by the model calculation incorporating ultrafast demagnetization. This fact indicates that thermal aspect of optical excitation is not responsible for the PEPM in (Ga,Mn)As, at least in the regime of weak excitation.

We now come to analysis of signals in ultrafast time region (Figs. 1(c) and 1(d))

based on autocorrelation function [29]. Let us define the delay time $t_{delay}$ in P-P experiment as the time interval between a pump pulse $I_{pu}(t) = I_{pu}^0 G(t) = |E_{pu}^0(t)|^2$ and a probe pulse $I_{pr}(t - t_{delay}) = I_{pr}^0 G(t - t_{delay}) = |E_{pr}^0(t - t_{delay})|^2$, where $G(t)$ is the Gaussian function. Optical response at a sample surface $\Delta r(t)$, induced by electric fields of both pulses, $E_{pu}(t) = E_{pu}^0(t)\exp(-i\omega t)$ and $E_{pr}(t) = E_{pr}^0(t)\exp(-i\omega t)$, is expressed as follows [29]:

$$\Delta r(t) = \Delta r^0 \int_{-\infty}^{t} f(t-\tau)|E_{pu}(\tau) + E_{pr}(\tau)|^2 d\tau \qquad (2)$$

Here, the sample response function is approximated by $f(t) = \sum_i f_i \exp(-t/T_i)$ with $T_i$ being the lifetime of excitation. Consequently, the intensity of reflected probe light is expressed by Eq. 3:

$$S^{(1)}(t) = \int_{-\infty}^{\infty} |\{r^0 + \Delta r(\tau')\} E_{pr}(\tau')|^2 d\tau' \qquad (3)$$

The incoherent process can be expressed by omitting the interference terms $E_{pu}^* E_{pr}$ and $E_{pu} E_{pr}^*$ in Eqs. 2 and 3, which leads us to Eq. 4:

$$S^{(2)}(t) = \int_{-\infty}^{\infty} \{r^0 + \Delta r(\tau')\} I_{pr}(\tau') d\tau' \qquad (4)$$

Closed symbols in Fig. 2(b) are experimental temporal profiles between −1 and 2 ps obtained for $h\nu = 1.57$ eV for five different excitation intensities up to 10 $\mu J/cm^2$. Solid lines in Fig. 2(b) are the fits to the experimental data using Eq. 4 with the standard deviation in G($t$), $\sigma$ = 48.9 fs, $I_{pr}^0$ = 84 nJ/cm$^2$, and the adjustable parameter $\Delta r^0/r^0$ = 0.01. No correction due to the negative demagnetization signals was necessary. From those fits, we find that the response function $f(t)$ commonly consists of two exponential decay components: the primary one with $T_1$ = 180 fs and the secondary one with $T_2$ = 900 fs. The origin of the $T_1$ value could be attributed to spectral hole burning and carrier thermalization, whereas that of the $T_2$ value to the conduction band carrier lifetime [27]. From these analyses, it is clear that the spike-like component can be understood quantitatively in terms of a fast change in reflectivity due to photo-generation of carriers and their relaxation.

Calculation with Eq. 3 with $I_{pu}^0$ = 1.7 $\mu$J/cm$^2$, $I_{pr}^0$ = 84 nJ/cm$^2$, and $\sigma$ = 38.6 fs, together with the adjustable parameter $\Delta r^0/r^0$ = 0.01 and $f(t)$ = exp(-$t$/40 fs) successfully results in rapidly oscillating signals similar to the experimental data at $h\nu$ = 1.44 eV (Fig. 1(d) and inset). As shown in Fig. 2(c), the amplitude of rapidly oscillating signals increases with increasing the excitation fluence, all of which are well reproduced by the autocorrelation model with the time constant of 40 fs (not shown in the figure). These facts suggest that the lifetime of photo-generated carriers, regardless of those in the conduction or valence bands, is getting extremely short, being less than the time scale of intraband thermalization. As for carriers in the conduction band, ultrafast trapping by arsenic antisite defects is one of the most plausible processes [32].

Before closing, we address a microscopic mechanism which triggers PEPM. Experiments with P-P energy of $h\nu = 1.38 - 1.51$ eV have revealed that the lifetime of carriers is extremely short (< 40 fs), whereas PEPM is still active. As a plausible scenario which gives rise to a non-thermal change in magnetic anisotropy, we suggest the photo-ionization process, $Mn^{2+}$ ($3d^5$) → $Mn^{3+}$ ($3d^4$) (Fig. 2(d)): namely, excitation of electrons from the $Mn^{2+}$ ($3d^5$) states to the conduction band of the host GaAs. Recall that incorporation of large amount of Mn ions does not result in the continuous shift of band edges, but results in the development of relatively broad (50 ~ 100 meV) Mn impurity band [33,34], as found by x-ray photo-emission spectroscopy [35] and scanning tunneling microscopy [36]. Furthermore, angular momentum can emerge after the excitation with linearly polarized light, since $g = 2.0$ for the $Mn^{2+}$ ($3d^5$) and $g = 2.8$ for $Mn^{3+}$ ($3d^4$) [37]. As a whole, the excitation of the Mn band with photons of less than the GaAs band gap affects the direction of $H_{eff}$. The slow raise/decline rate of $H_{eff}$ rotation (a broken line in Fig. 1(b) and [14]) can probably be associated with carrier repopulation process within the Mn band in which the rate determining process is the kinetics of electrons trapped in defect states. Our newly proposed scenario should be examined also in samples with higher Mn contents [38]. Microscopic mechanism of the out-of-plane rotation of $H_{eff}$, which suggests the breakdown of symmetry along the axis normal, is another interesting subject to be pursued. Concentration gradient of $Mn^{3+}$ may be a reasonable starting point to consider this problem.

In summary, we have established that PEPM in (Ga,Mn)As is not triggered by ultrafast demagnetization. In the view of optical excitation, (Ga,Mn)As can be regarded as an insulator, at least in the regime of weak excitation. We have showed furthermore that PEPM can be triggered by below-gap excitation which most likely involves photo-ionization of the $Mn^{2+}$ impurity states. The $Mn^{2+}$ states which are manifested by our experiments may be relevant to those discussed in the study of transport in (Ga,Mn)As-based resonant tunneling structure [39].

We acknowledge partial supports from Advanced Photon Science Alliance Project from MEXT and Grant-in- Aid for Scientific Research (No. 22226002) from JSPS.

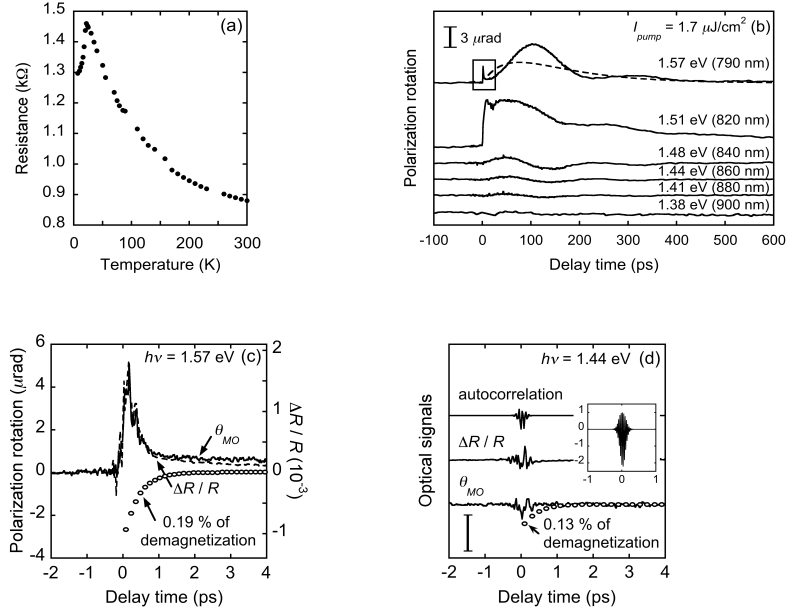

Fig.1. (a) Temperature dependence of resistance in (Ga,Mn)As sample. (b) Various temporal MO profiles obtained in sub-ns time region with six different P-P photon energies. Profile shown by a dashed line represents that of the rotated $H_{eff}$ towards out-of-plane direction. Spike-like component appearing for excitation with $h\nu \geqq 1.51$ eV is emphasized by a rectangle frame. (c) MO ($\theta_{MO}$, solid line) and DR ($\Delta R / R$, broken line) profiles obtained at $h\nu =$ 1.57 eV and $I_{pump} = 1.7$ μJ/cm$^2$. Theoretical profile due to ultrafast demagnetization is represented by open circles. (d) MO ($\theta_{MO}$) and DR ($\Delta R / R$) profiles obtained by excitation at $h\nu$ = 1.44 eV and $I_{pump} = 1.7$ μJ/cm$^2$, together with theoretical DR profile calculated based on autocorrelation function (26.67 fs step). Those curves are vertically shifted for graphic clarity. Scale bar represents polarization rotation of 2 μrad and reflectivity change of $5 \times 10^{-4}$. Theoretical profile due to ultrafast demagnetization is also shown by open circles. Inset shows calculated DR profile computed with a step of 2.667 fs. Here, intervals between tick marks on time (horizontal) and $\Delta R / R$ (vertical) axes are represented in the unit of ps and $10^{-4}$, respectively.

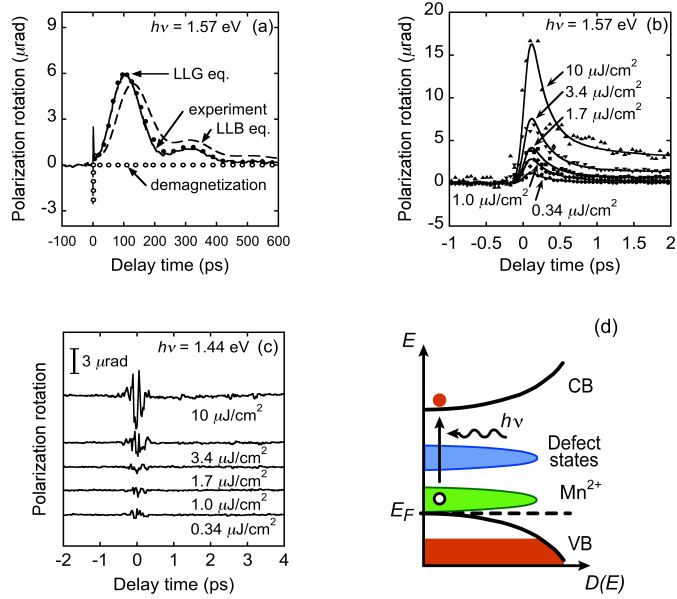

Fig.2. (a) Model calculation of MO profile with LLB (dashed line) and LLG (dots) equations. Open circles represent the simulated profile of ultrafast demagnetization. MO profile obtained experimentally at $h\nu$ = 1.57 eV and $I_{pump}$ =1.7 μJ/cm$^2$ is also shown together. (b) MO profiles obtained at $h\nu$ = 1.57 eV in ultrashort time region for five different pump fluences, 0.34 (circles), 1.0 (diamonds), 1.7 (squares), 3.4 (bottom-up triangles), and 10 μJ/cm$^2$ (bottom-down triangles). Lines are fit to each experimental data with autocorrelation function represented by Eq. (4). (c) MO profiles obtained at $h\nu$ = 1.44 eV in ultrashort time region for five different pump fluences. (d) Schematic illustration representing non-thermal mechanism of PEPM involving photo-ionization of Mn$^{2+}$ impurity band (colored in green for on-line version) in (Ga,Mn)As. CB and VB stand for the conduction and valence bands of host GaAs, respectively. In the valence band, white region denotes states occupied by holes. Position of the Fermi level is drawn in between Mn impurity band and VB.